# Radiation pressure on a submerged absorptive partial reflector deduced from the Doppler shift

Masud Mansuripur[†] and Armis R. Zakharian[‡]

[†]College of Optical Sciences, The University of Arizona, Tucson AZ 85721
[‡]Corning Incorporated, Science and Technology Division, Corning, New York 14831

**Abstract**. When a light pulse is reflected from a mirror, energy and momentum are exchanged between the electromagnetic field and the material medium. The resulting change in the energy of the reflected photons is directly related to their Doppler shift arising from the change in the state of motion of the mirror. Similarly, the Doppler shift of photons that enter an absorber is intimately tied to the kinetic energy and momentum acquired by the absorber in its interaction with the incident light. The argument from the Doppler shift yields expressions for the exchanged energy and momentum that are identical with those obtained from Maxwell's equations and the Lorentz law of force, despite the fact that the physical bases of the two methods are fundamentally different. Here we apply the Doppler shift argument to a submerged partial reflector (one that absorbs a fraction of the incident light), deducing in the process the magnitude of the photon momentum within the submerging medium. We also discuss the case of the submerging medium having a negative refractive index, and show the absence of the so-called "reversed" Doppler shift when the reflector is detached from the negative-index medium.

**1. Introduction**. The question of the momentum of light inside material media has been debated for over a century [1-22]. In Max Abraham's formulation [3], the linear momentum density is given by $\boldsymbol{E} \times \boldsymbol{H}/c^2$, while in Hermann Minkowski's theory [4] the same entity appears as $\boldsymbol{D} \times \boldsymbol{B}$. Here $\boldsymbol{E}$ is the electric field, $\boldsymbol{H}$ the magnetic field, $\boldsymbol{D}$ the displacement, $\boldsymbol{B}$ the magnetic induction, and $c$ the speed of light in vacuum [23,24]. If one considers a long, large-diameter light pulse of angular frequency $\omega_o$ traveling inside a transparent dielectric of refractive index $n_o$, the energy of the pulse may be normalized by $\hbar\omega_o$ to yield its photon content; here $\hbar$ is the reduced Planck constant. The momentum of the pulse divided by the number of photons then yields the expressions $\hbar\omega_o/(n_o c)$ and $n_o\hbar\omega_o/c$ for the Abraham and Minkowski photon momenta, respectively.

In recent years, a consensus seems to have emerged that the Abraham formula gives the correct electromagnetic (EM) momentum [6,10,11,12,21,22], but the interpretation of the Minkowski momentum remains unsettled. Barnett [19], for example, has argued that Minkowski's is the canonical momentum, while our own studies indicate that a consistent application of the Lorentz law of force would require the total (i.e., electromagnetic plus mechanical) photon momentum to have the arithmetic mean of the Abraham and Minkowski values, i.e., ½$(n_o+n_o^{-1})\hbar\omega_o/c$ [12,20,25-27].

A powerful argument in favor of Minkowski's formulation has been based on the Doppler shift phenomenon, originally advanced by Milonni et al [28,29], and subsequently used by Barnett [19] in support of his contention that the Minkowski momentum is the canonical momentum of the photon. In the present paper, we analyze the reflection of light from a partially-absorbing mirror submerged in a dielectric medium and, using the Doppler-shift argument, derive a general expression for the mechanical momentum imparted to the mirror. We then show that our analysis is fully consistent with the results obtained by a direct application of

Maxwell's equations and the Lorentz law of force to the submerged mirror. The conclusion is that the Doppler-shift argument does not contradict our earlier results, and that the photon momentum inside a dielectric medium does indeed have the mean value of the Abraham and Minkowski momenta. Similar conclusions were reached in a recent paper in which we analyzed the Doppler shift from a *perfect* reflector submerged in a transparent dielectric [27]; however, the case of a submerged *partial* reflector, which is the subject of the present paper, is much more general and covers the entire range of submerged objects from perfect absorbers to perfect reflectors.

The Doppler-shift analysis requires the motion of the absorbing/reflecting object within a submerging liquid of refractive index $n_o$. A stationary object acquires some momentum upon interaction with an incident EM wave, and begins to move with a nonzero velocity $V$, even though $V$ may be exceedingly small for a massive object – an object whose mass $m$ is much greater than the energy of the light pulse divided by $c^2$. When calculating the Lorentz force exerted by the EM field on the object, we generally assume that $m$ is large enough that any movement of the object as well as its acquired kinetic energy $\tfrac{1}{2}mV^2$ could be neglected, even though its acquired momentum $p=mV$ cannot be ignored. The validity of this assertion should be obvious considering that the kinetic energy, $p^2/(2m)$, goes to zero when $m\to\infty$, with $p$ remaining finite. With the Doppler-shift argument, however, the entire analysis revolves around the values of $V$ and $mV$ and $\tfrac{1}{2}mV^2$, despite the fact that they may be imperceptibly, even immeasurably, small. We will see that the Doppler-shift argument enables one to accurately determine the momentum acquired by the partial reflector, without ever having to explicitly evaluate the Doppler shift, the velocity $V$ of the reflector, or its kinetic energy $\tfrac{1}{2}mV^2$.

When an object moves inside a submerging liquid, there exist several possibilities as to how the liquid reacts to the motion. If the liquid is compressible (and stretchable), we must follow the propagation of elastic waves within the liquid and account for the concomitant changes of the refractive index. If the liquid is incompressible, then some sort of flow must be induced to fill the empty spaces created by the displacement of the object, or to replace the liquid that is dragged along. Allowing the entire system of object-plus-liquid to move together is another option, but this does not help with the question of the photon momentum inside the liquid, as the relative motion between the two is essential for the purpose. We handle the problem of motion within the liquid by surrounding the object with a small air-gap, within which the object can move freely, without disturbing the liquid. The gap between the object and its liquid host could be made as small as desired, since one can make the object's movement arbitrarily small by assigning it a sufficiently large mass. As far as EM fields and forces are concerned, perturbations caused by the introduction of an air-gap become negligible if the gap-width is orders of magnitude smaller than the optical wavelength. It is thus clear that, by introducing the artificial device of a tiny air-gap that surrounds the object, one can enormously simplify the problem without modifying its essential physics. In fact, with an air-gap isolating the object from the host medium, it is no longer necessary to assume that the host is in liquid form; the host could as well be replaced by a solid, transparent dielectric that has the same optical constants as the actual liquid host.

The paper is organized as follows. In Section 2 we describe the geometry of the system under consideration, and derive expressions for the absorbed as well as reflected EM fields in the presence of a submerged partial reflector. Section 3 is devoted to a calculation of the Doppler shifts of the absorbed as well as reflected light when the submerged absorber/reflector moves at a constant velocity within its "cocoon," closely surrounded by, yet not in contact with, the host dielectric. Numerical solutions of Maxwell's equations are then presented in Section 4, which confirm the theoretical results of the preceding section. The Doppler shifts associated with the absorbed and reflected beams uniquely specify the mechanical momentum imparted to the mirror in the process of absorption/reflection, as demonstrated in Section 5. In Section 6 we show that the momentum derived from the Doppler-shift argument is identical with that obtained from a direct calculation of the Lorentz force acting on the mirror material. The special cases examined in subsequent sections shed light on the nature of the photon momentum inside a dielectric host and its relation with the mechanical momentum imparted to the submerged partial reflector. The case of a negative-index submerging medium is briefly discussed in Section 11. Final thoughts and concluding remarks form the subject of the closing section.

**2. System description**. Consider a transparent dielectric slab of refractive index $n_o$, separated by an air-gap of width $d$ from an absorber/reflector of refractive index $n_1+i\kappa_1$, as shown in Fig. 1. A monochromatic, linearly-polarized plane-wave of frequency $\omega_o$ enters the dielectric slab and interacts with the partial reflector. The vacuum wavelength of the incident beam is $\lambda_o=2\pi c/\omega_o$, where $c$ is the speed of light in vacuum. The front facet of the slab is antireflection coated to avoid interference between the incident and reflected beams within the slab. Inside the glass slab at $z=0^-$, the (complex) $E$-field amplitudes of the incident and reflected beams are denoted by $E_o$ and $E_r$, respectively. The right-propagating plane-wave in the gap region has amplitude $E_g$ at $z=0^+$, while the $E$-field amplitude immediately inside the (imperfect) mirror at $z=d^+$ is denoted by $E_m$.

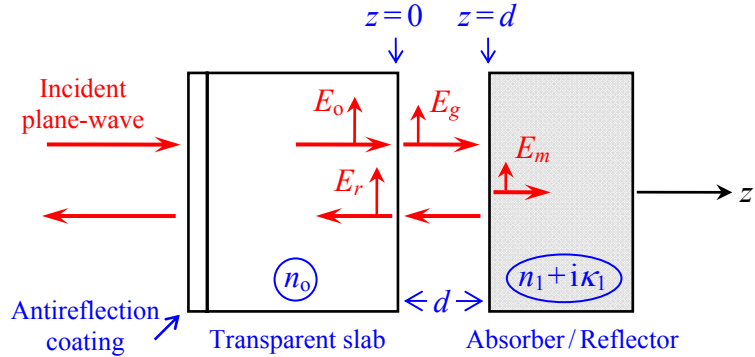

**Fig. 1** (Color online). A linearly-polarized, monochromatic plane-wave of frequency $\omega_o$ enters a transparent dielectric slab of refractive index $n_o$, whose front facet is antireflection coated. An airgap of width $d$ separates the slab from a partial reflector of refractive index $n_1+i\kappa_1$. The $E$-field amplitudes of the incident and reflected beams at the rear-facet of the transparent slab ($z=0^-$) are denoted by $E_o$ and $E_r$, respectively. There reside two counter-propagating plane-waves in the air gap, of which the forward-propagating wave has amplitude $E_g$ at $z=0^+$. The plane-wave that enters the partial reflector (and is eventually extinguished due to the nonzero absorption coefficient $\kappa_1$ of this medium) has amplitude $E_m$ at $z=d^+$.

Considering that the Fresnel reflection coefficients at the air-glass interface and at the air-mirror interface are $\rho_o=(1-n_o)/(1+n_o)$ and $\rho_1=(1-n_1-i\kappa_1)/(1+n_1+i\kappa_1)$, respectively, that the transmission coefficient at the glass-air interface is $\tau_o=2n_o/(1+n_o)$, and that the round-trip phase within the gap is $2\varphi_o=4\pi d/\lambda_o$, one may write the consistency equation at $z=0^+$ as follows:

$$E_g = \frac{2n_o}{(1+n_o)}E_o + \frac{(1-n_o)}{(1+n_o)}\frac{(1-n_1-i\kappa_1)}{(1+n_1+i\kappa_1)}\exp(i4\pi d/\lambda_o)E_g. \tag{1}$$

The above equation then yields

$$\frac{E_g}{E_o} = \frac{2n_o(n_1+i\kappa_1+1)\exp(-i\varphi_o)}{(n_o+1)(n_1+i\kappa_1+1)\exp(-i\varphi_o)-(n_o-1)(n_1+i\kappa_1-1)\exp(i\varphi_o)}. \quad (2)$$

Similarly, the $E$-field amplitude immediately inside the mirror, $E_m=(1+\rho_1)E_g\exp(i\varphi_o)$, may be written with the aid of Eq.(2) as follows:

$$\frac{E_m}{E_o} = \frac{4n_o}{(n_o+1)(n_1+i\kappa_1+1)\exp(-i\varphi_o)-(n_o-1)(n_1+i\kappa_1-1)\exp(i\varphi_o)}. \quad (3)$$

The power per unit area transmitted into the mirror is given by the Poynting vector $\mathbf{S}(z=d^+) = \tfrac{1}{2}\mathrm{Re}(\mathbf{E}_m\times\mathbf{H}_m^*)$, where $H_m=(n_1+i\kappa_1)E_m/Z_o$, with $Z_o=\sqrt{\mu_o/\varepsilon_o}$ being the impedance of free space. Similarly the incident optical power density at $z=0^-$ is $S_o=\tfrac{1}{2}n_o|E_o|^2/Z_o$. The ratio of the transmitted to incident optical power is thus found to be

$$\frac{P_m}{P_o} = \frac{n_1|E_m|^2}{n_o|E_o|^2} = \frac{8n_o n_1}{(n_o^2-1)[2\kappa_1\sin 2\varphi_o-(n_1^2+\kappa_1^2-1)\cos 2\varphi_o]+(n_o^2+1)(n_1^2+\kappa_1^2+1)+4n_o n_1}. \quad (4)$$

Next, we analyze the reflected beam. The consistency equation at $z=0^-$ may be written as follows:

$$E_r = \frac{n_o-1}{n_o+1}E_o + \frac{2}{(n_o+1)}\times\frac{(1-n_1-i\kappa_1)}{(1+n_1+i\kappa_1)}\exp(i4\pi d/\lambda_o)E_g. \quad (5)$$

With the help of Eq.(2), the above equation yields

$$\frac{E_r}{E_o} = \frac{(n_o-1)(n_1+i\kappa_1+1)\exp(-i\varphi_o)-(n_o+1)(n_1+i\kappa_1-1)\exp(i\varphi_o)}{(n_o+1)(n_1+i\kappa_1+1)\exp(-i\varphi_o)-(n_o-1)(n_1+i\kappa_1-1)\exp(i\varphi_o)}. \quad (6)$$

The ratio of the reflected to incident power is thus given by

$$\frac{P_r}{P_o} = \frac{|E_r|^2}{|E_o|^2} = \frac{(n_o^2-1)[2\kappa_1\sin 2\varphi_o-(n_1^2+\kappa_1^2-1)\cos 2\varphi_o]+(n_o^2+1)(n_1^2+\kappa_1^2+1)-4n_o n_1}{(n_o^2-1)[2\kappa_1\sin 2\varphi_o-(n_1^2+\kappa_1^2-1)\cos 2\varphi_o]+(n_o^2+1)(n_1^2+\kappa_1^2+1)+4n_o n_1}. \quad (7)$$

Clearly, $P_r+P_m=P_o$, which confirms the conservation of energy. In the next section, we consider the consequences of the motion of the mirror by allowing the gap-width $d$ to be a linear function of time. We then use Eqs.(3) and (6) to derive the Doppler shift of the EM waves upon entrance into as well as reflection from the imperfect mirror.

**3. Doppler shift caused by a moving mirror**. Let the mirror move at a constant velocity $V$ along the $z$ direction. The light entering the mirror will be Doppler-shifted to the new frequency $\omega_m=\omega_o-\Delta\omega_m$. To find $\Delta\omega_m$ we need the phase of $E_m/E_o$ of Eq.(3), which is the imaginary part of $\ln(E_m/E_o)$. Setting $d=d_o+Vt$ and differentiating the phase of $E_m/E_o$ with respect to time, we find

$\Delta\omega_m = \mathrm{Im}[\partial\ln(E_m/E_o)/\partial t|_{t=0}]$

$$= \frac{2[(n_o n_1 + 1)(n_o + n_1) + n_o \kappa_1^2](\omega_o V/c)}{(n_o^2 - 1)[2\kappa_1 \sin 2\varphi_o - (n_1^2 + \kappa_1^2 - 1)\cos 2\varphi_o] + (n_o^2 + 1)(n_1^2 + \kappa_1^2 + 1) + 4n_o n_1}. \quad (8)$$

The above Doppler-shift formula is accurate provided that the gap-width $d$ and the velocity $V$ are not too large, so that the gap-width would remain more or less constant while the light travels a few times back and forth inside the gap until the field $E_m$ of Eq. (3) is stabilized. We expect Eq. (8) to retain its validity over a wide range of values of $V$, up to, but not including, relativistic speeds. The numerical solutions of Maxwell's equations presented in the following section confirm our expectations. For purposes of the present paper, however, where $d$ and $V$ are exceedingly small, Eq. (8) should be highly accurate.

The reflected light will also be Doppler-shifted to a new frequency $\omega_r = \omega_o - \Delta\omega_r$. As before, we find $\Delta\omega_r$ by setting $d = d_o + Vt$ and differentiating the phase of $E_r/E_o$ of Eq. (6) with respect to time. We find

$$\Delta\omega_r = \mathrm{Im}[\partial \ln(E_r/E_o)/\partial t |_{t=0}]$$

$$= \frac{2[(n_o n_1 + 1)(n_o + n_1) + n_o \kappa_1^2](\omega_o V/c)}{(n_o^2 - 1)[2\kappa_1 \sin 2\varphi_o - (n_1^2 + \kappa_1^2 - 1)\cos 2\varphi_o] + (n_o^2 + 1)(n_1^2 + \kappa_1^2 + 1) + 4n_o n_1}$$

$$- \frac{2[(n_o n_1 - 1)(n_o - n_1) - n_o \kappa_1^2](\omega_o V/c)}{(n_o^2 - 1)[2\kappa_1 \sin 2\varphi_o - (n_1^2 + \kappa_1^2 - 1)\cos 2\varphi_o] + (n_o^2 + 1)(n_1^2 + \kappa_1^2 + 1) - 4n_o n_1}. \quad (9)$$

To find the average (or effective) Doppler shift $\Delta\omega_{\mathrm{eff}}$, we multiply $\Delta\omega_m$ and $\Delta\omega_r$ by the probabilities of absorption ($P_m/P_o$) and reflection ($P_r/P_o$), respectively, then add the results to obtain

$$\Delta\omega_{\mathrm{eff}} = \frac{P_m \Delta\omega_m + P_r \Delta\omega_r}{P_o}$$

$$= \frac{4n_o(n_1^2 + \kappa_1^2 + 1)(\omega_o V/c)}{(n_o^2 - 1)[2\kappa_1 \sin 2\varphi_o - (n_1^2 + \kappa_1^2 - 1)\cos 2\varphi_o] + (n_o^2 + 1)(n_1^2 + \kappa_1^2 + 1) + 4n_o n_1}. \quad (10)$$

The above result, when multiplied by $\hbar$, yields the total reduction in the energy of a single photon in its interaction with the system of slab-plus-mirror. Needless to say, we are speaking here of the average behavior of a large number of photons contained in an incident light pulse. A single photon is either absorbed or reflected with the aforementioned probabilities, but, *on average*, the energy given up by each photon is going to be $\hbar\Delta\omega_{\mathrm{eff}}$.

**4. Computer simulations**. The Finite Difference Time Domain (FDTD) technique [30] is ideally suited for simulations of the Doppler shift. As an example, consider a short light pulse of duration $\tau$ and central wavelength $\lambda_o = 2\pi c/\omega_o$, propagating in a homogeneous medium of refractive index $n_o$ along the $z$-axis, as depicted in Fig. 2(a). A flat mirror, placed perpendicular to the $z$-axis, reflects the light pulse. For a stationary mirror, the spectrum of the reflected light may be readily computed by Fourier transforming its $E$- and $H$-field amplitudes at a fixed point along the $z$-axis, then evaluating the Poynting vector $\boldsymbol{S}(\omega) = \tfrac{1}{2}\mathrm{Re}[\boldsymbol{E}(\omega) \times \boldsymbol{H}^*(\omega)]$ at each frequency $\omega$. Shown in Fig. 2(b) are two such spectra for the case of $\tau = 40$ fs, $\lambda_o = 633$ nm, $n_o = 1$ (solid black curve) and $n_o = 1.33$ (solid blue curve). The reflector in both cases is a dielectric slab of refractive index $n_1=2$ contiguous with the host medium of refractive index $n_o$. The rear facet of the reflector makes contact with the perfectly-matched boundary layer, which acts as a perfect absorber, simulating a non-reflecting, open boundary condition.

To simulate the motion of the mirror along the positive $z$-axis, we continually reduced the refractive index $n_1$ of the mirror's front facet (a single sheet of pixels within the FDTD mesh) until it became equal to the refractive index $n_o$ of the host dielectric. At this point the second layer of the mirror becomes its new front facet, whose refractive index must then be gradually reduced toward that of the host dielectric. The process continues at a fixed rate, with successive layers of the mirror transformed into the material that forms the incidence medium. Meanwhile the incoming light pulse continues to strike the shifting surface of the mirror, reflecting off the host/mirror interface, and returning to the incidence medium, where it now propagates in the negative $z$ direction. (We mention in passing that numerical inaccuracies arise if the mirror velocity is too small compared to the speed $c$ of light in vacuum. However, for mirror velocities greater than about $10^{-4}c$, FDTD simulations yield highly accurate values for the Doppler shift.)

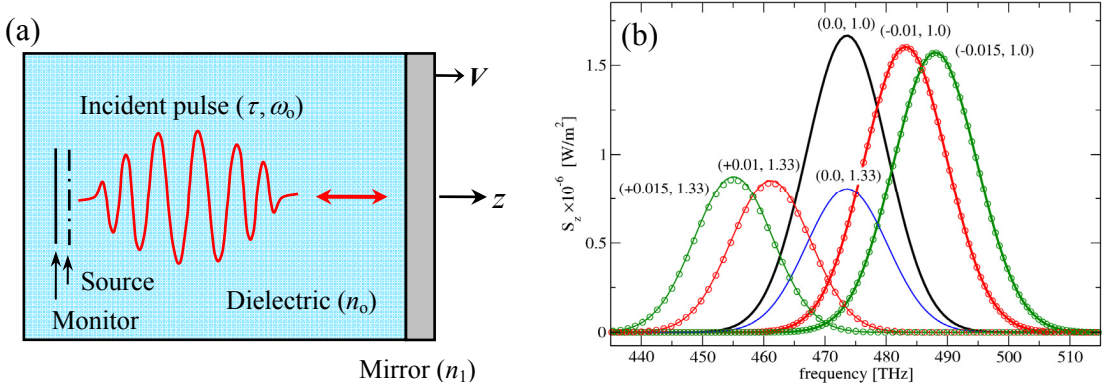

**Fig. 2** (Color online). (a) A 40 femtosecond Gaussian pulse having center wavelength $\lambda_o = 633$ nm is reflected from a moving mirror (velocity $=V$) immersed in a stationary dielectric medium of refractive index $n_o$. The mirror is a dielectric slab of refractive index $n_1 = 2.0$, whose rear facet is contiguous with the perfectly-matched boundary layer of the FDTD mesh. (b) Simulated spectra of the reflected pulse obtained with the FDTD method for different combinations of $V$ and $n_o$. The solid curve labeled $(V/c, n_o) = (0, 1.0)$ represents a stationary mirror in vacuum. The adjacent curve (solid red) labeled $(-0.01, 1.0)$ corresponds to a mirror moving in vacuum toward the source at $V = -0.01c$. The rightmost curve (solid green) is the reflected spectrum from a mirror moving in vacuum toward the source at $V = -0.015c$. The remaining three curves with reduced heights (solid blue, red, green) correspond to a mirror immersed in water ($n_o = 1.33$), moving away from the source at $V = 0$, $+0.01c$ and $+0.015c$, respectively. In each case, the superposed circles (red and green) are obtained by scaling the spectrum of the pulse reflected from a stationary mirror by the Doppler formula $\omega' = \omega(1 - n_o V/c)^2/(1 - V^2/c^2)$ along the frequency axis, and by the Lorentz transformation of the Poynting vector along the vertical axis. The simulated spectra (solid) are seen to closely mimic the Doppler-shifted spectra (circles).

Figure 2(b) shows the reflected spectra for different mirror velocities. In the case where the incidence medium is vacuum (i.e., $n_o = 1$), the simulated mirror velocities are $V/c = -0.01$ and $-0.015$; the reflected spectra (solid red and solid green) are seen to be blue-shifted, in excellent agreement with the (relativistic) Doppler formula $\omega' = \omega(1 - n_o V/c)^2/(1 - V^2/c^2)$ (small circles). In the case where the incidence medium is water (i.e., $n_o = 1.33$), the simulated mirror velocities are $V/c = +0.01$ and $+0.015$; the reflected spectra (solid red and solid green) are seen to be red-shifted, again in very good agreement with the aforementioned Doppler formula (small circles).

Note in the above simulations that it has been possible to successively replace layers of the mirror material with those of the host medium (or vice versa), thus simulating a smooth and uniform motion of the mirror without complications arising, for example, from laminar or turbulent flow, from induced density gradients, or from elastic wave generation and propagation within the host dielectric. Such problems, however, inevitably arise in an experimental setting, which would involve motion of the host liquid, interactions between the light pulse and the moving liquid (including exchanges of energy and momentum), and inaccuracies in the applicable Doppler shift formula caused by departures from uniformity of the liquid as well as its motion. To avoid such complications, we have chosen in this paper to analyze the case of a moving mirror detached from a solid, rigid dielectric block. The next simulation, therefore, addresses the case of a dielectric slab of refractive index $n_o=1.5$, separated from a moving mirror by an initial air-gap $d_o=100$ nm, as shown in Fig. 3(a). The results of our FDTD simulations appear in Fig. 3(b). Here a collimated beam of light, emanating from a continuous wave (cw) source located in free space, travels toward an antireflection-coated dielectric slab and, upon reflection from the mirror, returns to a monitor (also located in free space). In Fig. 3(b), the green (light gray) curve shows the time dependence of the $E$-field amplitude of the beam reflected from a mirror moving at $V=10^{-4}c$, while the blue (black) curve corresponds to $V=0.01c$.

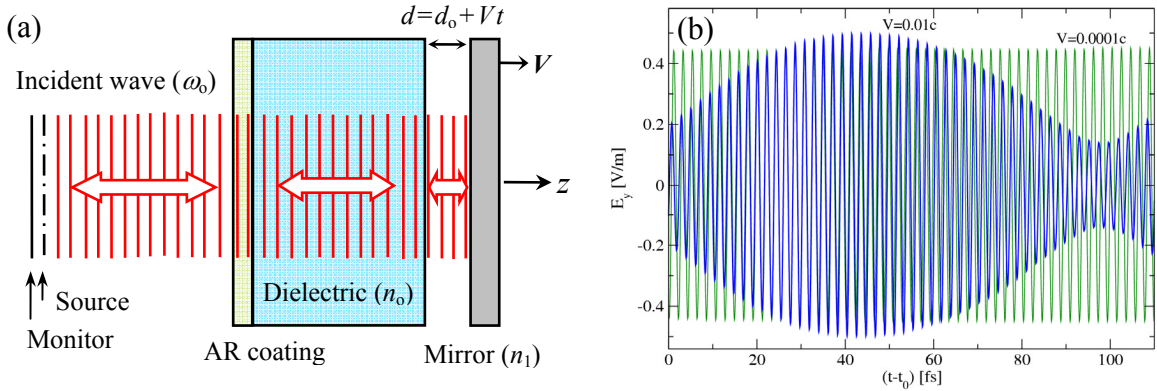

**Fig. 3** (Color online). (a) Monochromatic plane-wave is reflected from a flat mirror located behind a glass slab of refractive index $n_o=1.5$. The front facet of the slab is antireflection coated, the initial air-gap between the mirror and the slab is $d_o=100$ nm, and the mirror moves at a constant velocity $V$ away from the slab. In these simulations, the mirror is a dielectric slab of refractive index $n_1=2.0$, whose rear facet is contiguous with the perfectly-matched boundary layer of the FDTD mesh. (b) Time dependence of the $E$-field of the reflected beam monitored at a point slightly to the left of the light source. The green (light gray) curve corresponds to $V=10^{-4}c$, while the blue (black) curve represents the case of $V=0.01c$. The fields are captured starting at time $t_o$ (set to 90 fs in all simulations), to allow the transients initiated at the source startup to leave the computational domain.

At $V=10^{-4}c$, since the gap widens by only 3.3 nm during the 110 fs observation time, the effect of this change in the gap-width on the magnitude of the reflected $E$-field is negligible, which is why the dashed (green) curve's envelope is essentially constant. In contrast, at $V=0.01c$, the gap widens by 330 nm during the same time interval. The reflected $E$-field thus exhibits a significant amplitude modulation in consequence of this change in the gap-width ($V\Delta t > \lambda_o/2$), which is clearly visible in the envelope of the solid (blue) curve.

To monitor the phase of the reflected beam, we ignored the variations in the $E$-field magnitude and relied solely on the zero-crossings of the reflected field. The Doppler shift was thus estimated from these zero-crossings for various velocities of the mirror. For the system depicted in Fig. 3(a), the computed Doppler shifts $\Delta\omega_r$ versus time are plotted as solid curves in Figs. 4(a-c) for $V = 10^{-4}c$, $V = 3\times 10^{-3}c$, and $V = 10^{-2}c$, respectively. The observed variations of $\Delta\omega_r$ with time are a consequence of the fact that the gap-width $d$ continually increases with time during the FDTD simulations. For comparison, we have also plotted in Fig. 4 the theoretical values of $\Delta\omega_r$ (dashed curves) derived in Eq. (9). While at large (but not too large) mirror velocities the agreement between theory and simulation is superb, in the case of Fig. 4(a) corresponding to $V = 10^{-4}c$, there is a slight systematic error that is probably attributable to the limits of numerical accuracy attainable in this type of simulation. Even in this case, however, the difference between theoretical and simulated values of $\Delta\omega_r$ is well below 1%.

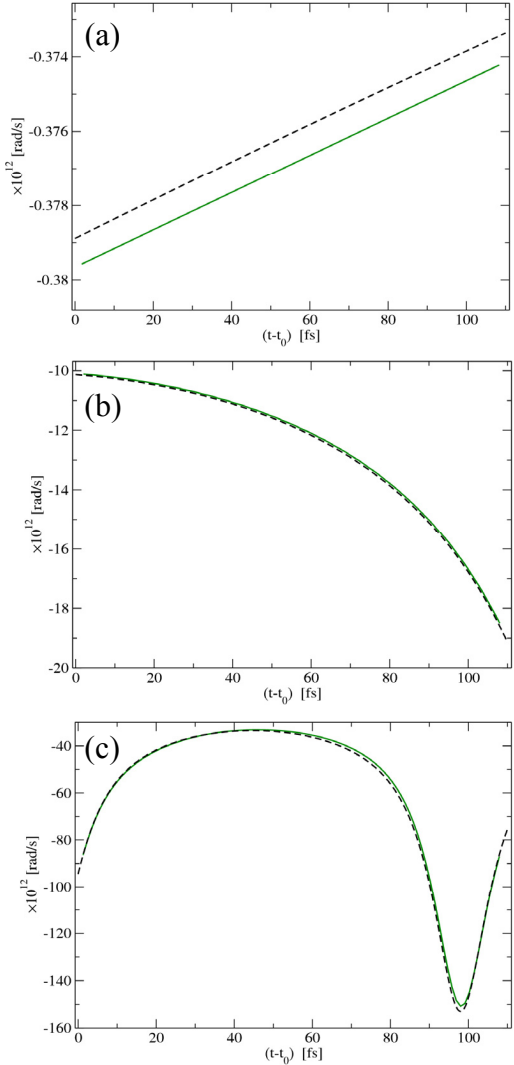

**Fig. 4** (Color online). Computed Doppler shift $\Delta\omega_r$ versus time for the reflected beam in the system of Fig. 3(a). From (a) to (c) the mirror velocity is $V = 0.0001c$, $0.003c$, and $0.01c$, respectively. In each case, the dashed curve is the theoretical estimate of $\Delta\omega_r$ given by Eq. (9). The theoretical estimate has been shifted forward in time to account for the delay caused by the propagation of the reflected beam from the rear facet of the glass slab to the observation point. In (c) the gap between the slab and the mirror widens by $\Delta d = 316.5$ nm between $t = t_o$ and $t = t_o + 105.5$ fs. This increase by $\lambda_o/2$ of the gap-width has no effect on the optical properties of the system, causing the plotted Doppler shift to repeat itself at regular intervals of $\Delta t = 105.5$ fs.

Note that the remarkable agreement between our FDTD simulation results depicted in Fig. 4 and the theoretical estimate of $\Delta\omega_r$ obtained in Eq. (9) confirms the validity of the assumptions that led to Eq. (9). In particular, we infer that the time needed for the stabilization of fields within the widening gap between the slab and the mirror is sufficiently short compared to the time it takes the mirror to move by a significant fraction of a wavelength. According to Fig. 4(c), this requirement seems to have been largely satisfied even at the substantial velocity of $V = 10^{-2}c$.

**5. Momentum imparted to the mirror**. With reference to the system of Fig. 1, suppose that a single incident photon imparts a mechanical momentum $p = mV$ to an initially stationary mirror ($p$ and $V$ are along the $z$-axis). We assume that the mass $m$ of the mirror is large enough that its acquired velocity $V$ is exceedingly small and that, therefore, the Doppler shift calculations of Sec. 3 are accurate. It is also necessary to assume a much larger mass $M$ for the glass slab than for the mirror, so that, irrespective of any momentum that the slab might pick up, its velocity as well as its kinetic energy will be negligible compared to the velocity and kinetic energy of the mirror.

Since the mirror is stationary at first, but acquires a velocity $V$ at the end of the reflection/absorption process, the effective Doppler shift of Eq. (10) should be associated with its average velocity $\tfrac{1}{2}V$. (Note that the constant radiation pressure of a rectangular pulse on the mirror gives it a constant acceleration in accordance with Newton's second law. The velocity thus increases linearly from zero to $V$, while the gap-width $d$ remains essentially constant, because the acquired velocity $V$ of a massive mirror is exceedingly small. The average Doppler shift is therefore associated with an average velocity $\tfrac{1}{2}V$ and a constant gap-width $d$.) The reduction of the photon energy $\hbar\Delta\omega_{\text{eff}}$ corresponding to a mirror velocity $\tfrac{1}{2}V$ must then be equated with the kinetic energy $\tfrac{1}{2}mV^2$ gained by the mirror. With $V$ dropping out from both sides of the equation, the momentum (per photon) taken up by the mirror in the reflection/absorption process must be

$$p = mV = \frac{4n_o(n_1^2 + \kappa_1^2 + 1)(\hbar\omega_o/c)}{(n_o^2 - 1)[2\kappa_1 \sin 2\varphi_o - (n_1^2 + \kappa_1^2 - 1)\cos 2\varphi_o] + (n_o^2 + 1)(n_1^2 + \kappa_1^2 + 1) + 4n_o n_1}. \quad (11a)$$

The above equation is the main result of the present paper; it relates the momentum per photon acquired by a submerged, partially-reflecting mirror to the system parameters $n_o$, $n_1$, $\kappa_1$, $\varphi_o$, and to the vacuum momentum $\hbar\omega_o/c$ of the photon. Strictly speaking, we should set $\varphi_o = 0$ for a submerged mirror, as the width $d$ of the gap between the mirror and its surrounding liquid would be negligible. However, the system of Fig. 1 is much more amenable to experimental verification than a system consisting of a mirror submerged in an actual liquid. For this reason, we shall continue to work with Eq. (11a) and to treat $\varphi_o$ as an adjustable parameter, with the understanding that, if the experiment is conducted with a liquid dielectric, then $\varphi_o$ must be set to zero, in which case we will have

$$p = \frac{2n_o(n_1^2 + \kappa_1^2 + 1)(\hbar\omega_o/c)}{(n_o + n_1)^2 + \kappa_1^2}. \quad (11b)$$

Before embarking on a description of the various features of Eq. (11), we show in the following section its precise agreement with the results of direct calculations using the Lorentz force law.

**6. Direct calculation of the mirror momentum**. The time-averaged Poynting vector for the forward-propagating beam in the gap (amplitude = $E_g$) is $\tfrac{1}{2}|E_g|^2/Z_o$. The momentum density in vacuum is given by the Poynting vector divided by $c^2$. Multiplying this density by the cross-sectional area $A$ and by the pulse length $cT$, where $T$ is the duration of a light pulse, we find the forward-propagating momentum to be $\tfrac{1}{2}\varepsilon_o|E_g|^2 AT$. Multiplication by the mirror reflectivity $|\rho_1|^2 = [(1-n_1)^2 + \kappa_1^2]/[(1+n_1)^2 + \kappa_1^2]$ yields the reflected momentum, and the total momentum transferred to the mirror is thus given by

$$\tfrac{1}{2}\varepsilon_o|E_g|^2 AT(1+|\rho_1|^2) = \varepsilon_o|E_g|^2 AT(n_1^2 + \kappa_1^2 + 1)/[(n_1+1)^2 + \kappa_1^2]. \tag{12}$$

The incident pulse energy is $\tfrac{1}{2}n_o|E_o|^2 AT/Z_o$. Normalizing the transferred momentum by the pulse energy, then multiplying by $\hbar\omega$ yields the transferred momentum per photon as follows:

$$\frac{2|E_g|^2(n_1^2 + \kappa_1^2 + 1)(\hbar\omega/c)}{n_o|E_o|^2[(n_1+1)^2 + \kappa_1^2]} =$$

$$\frac{4n_o(n_1^2 + \kappa_1^2 + 1)(\hbar\omega/c)}{(n_o^2 - 1)[2\kappa_1 \sin 2\varphi_o - (n_1^2 + \kappa_1^2 - 1)\cos 2\varphi_o] + (n_o^2 + 1)(n_1^2 + \kappa_1^2 + 1) + 4n_o n_1}. \tag{13}$$

This is the same result as obtained using the Doppler shift argument in the preceding section. The same exact result may also be derived by a direct calculation of the force on the mirror using the Lorentz force-density expression

$$(\partial \boldsymbol{P}/\partial t) \times \mu_o \boldsymbol{H} = \tfrac{1}{2}\mu_o\varepsilon_o \mathrm{Re}[-\mathrm{i}\omega_o(\varepsilon_1 - 1)\boldsymbol{E}_m(\boldsymbol{r},t)H_m^*(\boldsymbol{r},t)]\hat{\boldsymbol{z}}. \tag{14}$$

Here $\varepsilon_1 = (n_1 + \mathrm{i}\kappa_1)^2$ and

$$\boldsymbol{E}_m(z,t) = E_{mo}\hat{\boldsymbol{x}} \exp\{\mathrm{i}(\omega_o/c)[(n_1 + \mathrm{i}\kappa_1)(z-d) - ct]\}, \tag{15a}$$

$$\boldsymbol{H}_m(z,t) = (n_1 + \mathrm{i}\kappa_1)Z_o^{-1}E_{mo}\hat{\boldsymbol{y}} \exp\{\mathrm{i}(\omega_o/c)[(n_1 + \mathrm{i}\kappa_1)(z-d) - ct]\}. \tag{15b}$$

The force-density must be integrated along the penetration depth of the light into the mirror, i.e., from $z = d$ to $\infty$ to yield

$$F_z = \tfrac{1}{2}\mu_o\varepsilon_o\omega_o \mathrm{Im}[(\varepsilon_1 - 1)E_{mo}H_{mo}^*]\int_0^\infty \exp[-2(\omega_o/c)\kappa_1 z]\mathrm{d}z = \tfrac{1}{4}\varepsilon_o(n_1^2 + \kappa_1^2 + 1)|E_{mo}|^2. \tag{16}$$

The incident power per unit area is $\tfrac{1}{2}n_o|E_o|^2/Z_o$. We normalize the above force on the mirror by the incident optical power, then multiply by $\hbar\omega$ to determine the transferred momentum per photon as follows:

$$p = \frac{\tfrac{1}{4}\varepsilon_o(n_1^2 + \kappa_1^2 + 1)|E_{mo}|^2 \hbar\omega_o}{\tfrac{1}{2}n_o|E_o|^2/Z_o} = \tfrac{1}{2}n_o^{-1}(n_1^2 + \kappa_1^2 + 1)|E_{mo}/E_o|^2(\hbar\omega_o/c). \tag{17}$$

Substitution for $E_{mo}/E_o$ from Eq. (3) then yields the same expression as in Eq. (11a), which is also identical with the result obtained in Eq. (13). Note that the methods of calculation used in the present section are vastly different from those employed in Sec. 5. One method is based on the classical Maxwell's equations and the Lorentz law of force, the other relies on the quantum nature of light and the fact that individual photons of frequency $\omega_o$ carry the quantum of energy $\hbar\omega_{\text{o}}$. Despite these differences, the two methods predict exactly the same values for the momentum acquired by the mirror.

**7. Special case 1: Mirror in vacuum**. Let $n_{\text{o}}=1$, i.e., eliminate the glass slab from the system of Fig. 1. Equation (11a) will simplify as follows:

$$p = \frac{2(\hbar\omega_{\text{o}}/c)(n_1^2+\kappa_1^2+1)}{(n_1+1)^2+\kappa_1^2} = (1+|\rho_1|^2)(\hbar\omega_{\text{o}}/c). \tag{18}$$

Note that this result is independent of the gap-width $d$, depending only on the reflectivity $|\rho_1|^2$ of the mirror. The incident photon momentum, of course, is $\hbar\omega_{\text{o}}/c$, while the reflected momentum is $|\rho_1|^2\hbar\omega_{\text{o}}/c$. The fraction of energy that is not reflected at the surface will enter the mirror, and one must await its complete absorption before the mirror exhibits its full mechanical momentum $p$.

Another way of analyzing this problem is to note that $\Delta\omega_m = \omega_{\text{o}}V/c$ and $\Delta\omega_r = 2\omega_{\text{o}}V/c$ when $n_{\text{o}} = 1$. Therefore, an absorbed photon will transfer a momentum of $\hbar\omega_{\text{o}}/c$ to the mirror, while a reflected photon will transfer a momentum of $2\hbar\omega_{\text{o}}/c$. This is consistent with the fact that the photon momentum before arriving at the mirror is $\hbar\omega_{\text{o}}/c$ while that after reflection is $-\hbar\omega_{\text{o}}/c$. Taking into account the probabilities of absorption and reflection, namely, $1-|\rho_1|^2$ and $|\rho_1|^2$, respectively, we arrive once again at Eq. (18).

**8. Special case 2: Weakly-absorbing medium submerged in an index-matched fluid**. Let $d=0$, $n_1=n_{\text{o}}$, and $\kappa_1 \ll 1$. In this case the reflected power $P_r$ approaches zero as $\kappa_1 \to 0$, while the absorbed power $P_m$ approaches the incident power $P_{\text{o}}$. We will have

$$\frac{P_m}{P_{\text{o}}} = \frac{4n_{\text{o}}^2}{4n_{\text{o}}^2+\kappa_1^2} \xrightarrow[\kappa_1\to 0]{} 1. \tag{19a}$$

$$p = \frac{2n_{\text{o}}(n_{\text{o}}^2+\kappa_1^2+1)(\hbar\omega_{\text{o}}/c)}{4n_{\text{o}}^2+\kappa_1^2} \xrightarrow[\kappa_1\to 0]{} \tfrac{1}{2}(n_{\text{o}}+n_{\text{o}}^{-1})\hbar\omega_{\text{o}}/c. \tag{19b}$$

The momentum per photon transferred to the mirror thus has the arithmetic mean value of the Abraham and Minkowski momenta within the glass slab [12]. This, in fact, is the total (i.e., electromagnetic plus mechanical) momentum of a photon traveling inside a dielectric of refractive index $n_{\text{o}}$. Thus, in the limit of weak absorption ($\kappa_1 \to 0$), the photon is absorbed in the mirror with negligible probability of reflection. In this limit, the Doppler shift $\Delta\omega_m$ approaches $\tfrac{1}{2}(n_{\text{o}}+n_{\text{o}}^{-1})\omega_{\text{o}}V/c$, yielding the mirror momentum given by Eq. (19b). Here the photon momentum inside the dielectric slab is fully transferred to the absorber.

The fact that the mean value of the Abraham and Minkowski momenta is somewhat greater than the vacuum photon momentum can be accounted for once one recognizes that, upon entry into the glass slab, the photon exerts a pulling force on the antireflection layer; see Fig. 1. It is not difficult to confirm that the pull force on the antireflection coating precisely accounts for the excess momentum [12,20,31].

**9. Special case 3: Perfect reflector picking up the Minkowski momentum**. Let $d=0$, $n_{\text{o}}$ and $n_1$ arbitrary, and $\kappa_1 \gg 1$. In this case the reflected power approaches the incident power as $\kappa_1 \to \infty$. We will have

$$\frac{P_r}{P_o} = \frac{(n_o - n_1)^2 + \kappa_1^2}{(n_o + n_1)^2 + \kappa_1^2} \xrightarrow[\kappa_1 \to \infty]{} 1. \tag{20a}$$

$$p = \frac{2n_o(n_1^2 + \kappa_1^2 + 1)(\hbar\omega_o/c)}{(n_o + n_1)^2 + \kappa_1^2} \xrightarrow[\kappa_1 \to \infty]{} 2n_o\hbar\omega_o/c. \tag{20b}$$

The momentum transferred to the mirror is therefore twice the Minkowski momentum of the photon inside the dielectric slab. (Note that the mirror's reflection coefficient $\rho_1$ approaches $-1$ as $\kappa_1 \to \infty$; in other words, the phase of the Fresnel reflection coefficient in the limit of infinite $\kappa_1$ is 180°.) The fact that the momentum picked up by the mirror is $2n_o\hbar\omega_o/c$ does *not* imply that the photon inside the transparent slab carries the Minkowski momentum. The photon momentum before and after reflection is still $\frac{1}{2}(n_o + n_o^{-1})\hbar\omega_o/c$; what is different in this case is that, during the process of reflection, the incident and reflected EM fields interfere within the slab. In the overlap region, the fields exert a Lorentz force on the transparent medium which transfers a certain amount of mechanical momentum to the slab. Momentum balance is restored when this additional momentum is taken into account [12,32,33].

**10. Special case 4: Perfect reflector picking up the Abraham momentum**. Let $d = \frac{1}{4}\lambda_o$, $n_o$ and $n_1$ arbitrary, and $\kappa_1 \gg 1$. Once again the reflected power approaches the incident power as $\kappa_1 \to \infty$, but the momentum transferred to the mirror will be twice the Abraham momentum, as follows:

$$\frac{P_r}{P_o} = \frac{(n_o n_1 - 1)^2 + n_o^2 \kappa_1^2}{(n_o n_1 + 1)^2 + n_o^2 \kappa_1^2} \xrightarrow[\kappa_1 \to \infty]{} 1. \tag{21a}$$

$$p = \frac{2n_o(n_1^2 + \kappa_1^2 + 1)(\hbar\omega_o/c)}{(n_o n_1 + 1)^2 + n_o^2 \kappa_1^2} \xrightarrow[\kappa_1 \to \infty]{} 2\hbar\omega_o/(n_o c). \tag{21b}$$

As in the preceding example, the momentum transferred to the mirror in this case does not represent the photon momentum in the dielectric slab. Since the phase of the reflected beam at $z = 0^-$ is now 0° (as opposed to 180° in the previous example), the interference pattern that is set up between the incident and reflected fields is different. The Lorentz force acting on the dielectric slab in the region of interference thus differs from that in the preceding example. Once again, the photon momentum in the dielectric slab (both before and after reflection) has the mean value of the Abraham and Minkowski momenta, and the momentum balance is satisfied once the mechanical momentum picked up by the slab is taken into account [32].

**11. Negative-index medium (NIM)**. The case of a negative-index dielectric slab is of some interest, as it has been claimed that the Doppler shift in such media will be reversed [34,35]. We note that, whenever the permeability $\mu$ of the dielectric slab deviates from unity, our expressions for the absorbed and reflected optical power in Eqs.(4) and (7), as well as those for the Doppler shift in Eqs.(8) and (9), should be written in terms of the admittance $\eta_o = \sqrt{\varepsilon/\mu}$ of the material rather than in terms of its refractive index $n_o$. Since the admittance of a NIM is positive, and since the negative index $n_o$ does not appear anywhere in our equations, we conclude that the results of the preceding sections, with $n_o$ replaced by $\eta_o$, will remain valid for negative-index slabs as well.

In particular, the sign of the Doppler shift in the system under consideration here will *not* be reversed: As usual, the reflected light from a receding mirror, monitored either inside or outside the negative-index slab, will be red-shifted, while the reflected light from an approaching mirror will be blue-shifted. The arguments of the preceding sections thus apply equally to positive- and negative-index media, with the caveat that the admittance $\eta_o$ must take the place of $n_o$ in various formulas. For instance, the momentum per photon picked up by a perfect reflector immersed in a NIM could be anywhere between $2\hbar\omega_o/(\eta_o c)$ and $2\eta_o \hbar\omega_o/c$, depending on the phase of the Fresnel reflection coefficient of the mirror. As we have pointed out elsewhere [31], the actual photon momentum inside a NIM is $\frac{1}{2}(\eta_o + \eta_o^{-1})\hbar\omega_o/c$, of which the electromagnetic part is $\hbar\omega_o/(n_g c)$, the remainder being mechanical. Here $n_g$, the group refractive index, differs from $n_o$, the phase refractive index, as negative-index media must of necessity be dispersive.

We emphasize that our argument against the existence of a reversed Doppler shift in negative-index media applies only to cases where the reflector is detached from the NIM. When the moving mirror and the NIM maintain contact during the reflection process, as was the case in Veselago's original argument [34], we have no reason to doubt the possibility of existence of the reverse Doppler effect.

**12. Concluding remarks**. We have extended the analysis of a recent paper [27] for perfect mirrors submerged in a dielectric liquid to cover the case of imperfect mirrors that can absorb a fraction of the incident light. Deducing the radiation pressure on a submerged absorber/reflector from the Doppler shift (imperceptible as that shift may be), relies on a powerful new argument that is fundamentally different from our previous arguments based on classical electrodynamics. The reason for this difference is that the Doppler shift argument invokes the dependence of the photon energy on the oscillation frequency of the EM field ($\mathcal{E} = \hbar\omega$), whereas the energy content of light in classical electrodynamics is associated with the strength of the electric and magnetic fields. Despite this fundamental difference between the two methods of analysis, both yield identical results for the radiation pressure on submerged mirrors and absorbers.

Experimental results pertaining to photon momentum inside transparent dielectrics [36-39] have repeatedly confirmed the transfer of the Minkowski momentum to submerged mirrors, in agreement with the special case analysis of Section 9. It remains to conduct experiments with other types of reflectors, in particular those whose phase of the Fresnel reflection coefficient differs substantially from the conventional value of 180º, in order to confirm the possibility of transferring a wide range of momenta to submerged mirrors.